\documentclass{jnmp}
\usepackage{amsmath}
\usepackage{graphicx}
\JNMPnumberwithin{equation}{section}

\theoremstyle{definition}

\begin {document}

\thispagestyle{empty}
\FirstPageHead{*}{*}{2005}{\pageref{firstpage}--\pageref{lastpage}}{Article}
\copyrightnote{2005}{S~E Korenblit and V~V Semenov}
\Name{Massless Pseudo-scalar Fields and Solution of the Federbush Model}
\label {firstpage}
\Author{S~E Korenblit and V~V Semenov}

\Address {Irkutsk State University, Gagarin Blvd., 20, 664003 Irkutsk, Russia \\
E-mail: korenb@ic.isu.ru
}

\Date{Received Month *, 2005; Revised Month *, 2005;
Accepted Month *, 2005}

\begin {abstract}
\noindent 
The formal Heisenberg equations of the Federbush model are linearized 
and then are directly integrated applying the method of dynamical mappings.   
The fundamental role of two-dimensional free massless pseudo-scalar fields is 
revealed for this procedure together with their locality condition taken into 
account. Thus the better insight into solvability of this model is 
obtained together with the additional phase factor for its general solution, and  
the meaning of the Schwinger terms is elucidated.
\end {abstract}

\section {Introduction}  
Generalizing the methods, giving solutions in sufficiently
simple models one can hope to develop a technique to be effective for the
comparatively non-trivial and physically important problems \cite {whigh}. 
Examples of such partly or entirely solvable models are models with four-fermion 
interactions in two-dimensional
space-time and related to them non-linear bosonic models of the
sine-Gordon (SG) type. The relentless interest in these models is
due to the fact that their non-Abelian four-dimensional analogs
are more or less successfully used for the analysis and explanation 
of various non-perturbative effects in modern theory of strong interactions,
such as: description of the processes of quark hadronization and
the phenomenon of spontaneous symmetry breaking \cite {diakopetropolya}, 
\cite {polya}. They serve also as a ``testing ground'' for various 
non-perturbative methods \cite {belveamara}, \cite {belverodrig}, 
\cite {castralvarfring}, \cite {changrajar}, \cite {faberivano}. 
As a rule an operator solution for these models is obtained with the help of 
proper operator ansatz \cite {feder}, \cite {gomessilva1}, \cite {mande}, 
\cite {morchpierostroc2}, \cite {whigh} instead of consecutive integration 
the corresponding equations of motion.

Recently Faber, Ivanov in a series of papers (see e.g. \cite{faberivano}, 
\cite {faberivano2}), following Morchio et al. \cite {morchpierostroc2}, 
re-examined some ambiguities \cite {colem} of the Thirring model and elucidated 
the importance of existence of two-dimensional free massless scalar fields  for 
its solution.

One of the aims of the present work is to show a similar role of free 
massless \textit {pseudo-scalar} fields for a direct step-by-step integration 
of the Heisenberg equations (HEqs) of the Federbush model \cite {feder}, and to 
advocate as a general method the corresponding linearization procedure, 
successfully used to solve some non-relativistic and relativistic 
phenomenological models \cite {korentanae}, \cite {vallkorenlevia}. 
It is interesting to note that unlike free massless scalar field, the 
free massless pseudo-scalar field has a well-defined generating functional 
\cite {faberivano2} and appears in a definite sense as a gauge invariant object 
\cite {belverodrig}. 

Another goal is to understand the role of Schwinger terms for this 
model. We show that, in spite of a long story of this sloppy question 
\cite {Singh}, \cite {solomon}, \cite {vladimirov} the suggested linearization 
procedure allows naturally to adopt an idea of solution found in QED 
\cite {sokolov}.   

In order to linearize the HEqs we use the notion of physical fields 
\cite {blaso}, \cite {green2}, \cite {umeza}. Let us recall that the HEqs 
are formal relations between operators, as long as they are not defined on 
a corresponding vector space. This implies, that in order to give a physical 
meaning to the quantum-field-system description by means of the Heisenberg 
fields $\Psi(x)$, it is necessary to represent them in the space of physical 
states, that, in turn, necessitates to express them in terms of physical 
fields $\psi(x)$. The relationship between Heisenberg and physical fields 
is called dynamical mapping (DM), and may be formally expressed generally 
only as a weak equality: 
$\Psi(x)\stackrel {\rm w}{=}\Upsilon[\psi(x)]$. 
After quantization of the physical fields, satisfying some free field 
equations, one obtains the HEqs and DM in the normal-ordered form with 
respect to creation and annihilation operators of a chosen set of physical 
particles. This normal-ordered form in fact is equivalent in the sense 
of DM to their primary representations as a weak limit, but is, of course,
representation dependent \cite {blaso}.

Since dynamical HEqs contain time derivatives it is necessary to
choose initial conditions. There are two physically different
choices leading respectively to two essentially different sets of
physical fields: for asymptotic $in$-fields \cite {bjorkdrell}, 
\cite {umeza}, for $t\to-\infty$, one has 
$\lim\limits_{t\to-\infty}\Psi(x^1,t)\stackrel{\rm w}{=}
\Upsilon[\psi_{in}(x^1,-\infty)]$; alternatively for Schr{\"o}dinger 
fields \cite {shiro2} one has 
$\lim\limits_{t\to 0}\Psi(x^1,t)\stackrel{\rm w}{=}\Upsilon[\psi(x^1,0)]$ 
as $t\to 0$. The latter choice will be used here. It was advocated in our 
previous works \cite {korentanae}, \cite {vallkorenlevia} as some 
generalization of the usual interaction representation \cite {umeza} shown 
its usefulness on the bound-state problem, and in fact was already used for the 
construction of the solution of the Federbush model \cite {feder}.

The work is organized as follows. First in section 2 we discuss a linearization
procedure for HEqs. Then in section 3 we obtain their solutions in terms of 
Feynman functional integrals calculated in sections 4 and 5 by pure algebraic 
operator methods using various bosonization rules for the \textit {free fermion 
currents only} and the properties of free massless pseudo-scalar fields only.  
It should be emphasized in order to avoid some confusions, that, as is shown 
below, the bosonization rules in fact could not be applied here to the local 
Heisenberg currents directly, and that in two-dimension space-time both the used 
in section 4 bosonization rules: (\ref{4.2}) for the local free massive fermion 
current and (\ref{4.7}) for the local free massless fermion current, are immediate 
well-known consequence of conservation of both the vector currents \cite 
{gomessilva1}, \cite {whigh}. These two rules introduce however two different 
pseudo-scalar fields: the free massless field $\phi(x)$ in (\ref{4.7}) and the 
field $\Phi(x)$ in (\ref{4.2}) which in fact is the sine-Gordon field, as is 
finally checked in subsection 4.2. These fields certainly serve as building blocks 
for the construction of solution, nevertheless, are necessary but insufficient 
ingredients to obtain the explicit operator solution of the model, as is shown in 
sections 4 and 5. The role of Schwinger terms is also discussed in the Summary. 

\section{Step 1. Linearization of Heisenberg equations}
\resetfootnoterule

The model under consideration describes a parity non-conserving
local interaction of two different two-component fermion fields
$\Psi_{\xi}(x)$ in two-dimensional space-time with different masses $m_\xi$, 
$\xi=\pm 1$, and is defined by a Lagrangian density, bilinear in the two Noether 
vector currents 
$J_{\xi(\Psi)}^\nu(x)=\overline{\Psi}_\xi(x)\gamma^{\nu}\Psi_\xi(x)$.
Splitting the total Hamiltonian of the model into free and interaction parts 
as
\begin {gather}
H_{(\Psi)}(t)
=H_{0(\Psi)}(t)+H_{I(\Psi)}(t)
\label {2.1} \\
\equiv\int\limits^\infty_{-\infty}d{x^1}\sum\limits_{\xi=\pm 1}\left\{
\left(\Psi^{\dagger}_\xi(x)E_\xi(P^1)\Psi_\xi(x)\right)+\xi\pi\lambda
\epsilon_{\mu\nu}J^{\mu}_{\xi(\Psi)}(x)J^{\nu}_{-\xi(\Psi)}(x)\right\},
\label {2.2}
\end {gather}
and using the equal-time anticommutation relations for the
Heisenberg operators\footnote{Our conventions are: $x^\mu=(x^0,x^1)$, $x^0=t$,
$\hbar=c=1$, $\partial_{\mu}=(\partial_{0},\partial_{1})$, metric
tensor $g^{00}=-g^{11}=1$, antisymmetric tensor
$\epsilon^{01}=-\epsilon^{10}=1$, Dirac conjugate field
${\overline{\Psi}}_\xi(x)={\Psi}_\xi^\dagger(x){\gamma}^{0}$,
gamma matrices $\gamma^0=\sigma_1$, $\gamma^1=-i\sigma_2$,
$\gamma^5=\gamma^0\gamma^1=\sigma_3$, with $\sigma_i$ and $I$ as  
Pauli and unit matrices respectively, and operators
$P^1=-i\partial_1\equiv -i\partial/\partial x^1$,
$P^{1\dagger}=i\stackrel{\leftarrow}{\partial}_1$,
$E_\xi(P^1)=\gamma^5 P^1+\gamma^0m_\xi$.} 
\begin {gather}
\left\{\Psi_{\xi,\alpha }(x)\,,\Psi_{\xi^\prime,\beta}(y)\right\}
\biggr|_{x_0=y_0}\!\!\!=0,\;\;\;
\left\{\Psi_{\xi,\alpha }(x)\,,\Psi^{\dagger}_{\xi^\prime,\beta}(y)\right\}
\biggr|_{x_0=y_0}\!\!\!
=\delta_{\xi\xi^\prime}\delta_{\alpha\beta}\,\delta(x^1-y^1),
\label {2.3}
\end {gather}
we get the formal HEqs for the fields $\Psi_{\xi}(x)$
\begin {gather}
\!\!\!\!\!\!\!\! 
i\partial_0 \Psi_{\xi}(x)=\left[\Psi_{\xi}(x)\,,H_{(\Psi)}(t)\right]=
\left[ E_\xi(P^1)+{V}_{-\xi(\Psi)}(x)\right]{\Psi}_{\xi}(x), 
\label {2.4} \\
\!\!\!\!\!\!\!\!
\mbox{where }\, V_{-\xi(\Psi)}(x)=\xi 2\pi\lambda
\epsilon_{\mu\nu}{\gamma}^0{\gamma}^\mu J_{-\xi(\Psi)}^{\nu}(x)=
\xi 2\pi\lambda\left[\gamma^5J_{-\xi(\Psi)}^0(x)-IJ_{-\xi(\Psi)}^1(x)
\right],
\label {2.5}
\end {gather}
and since $\left[J^\nu_{-\xi(\Psi)}(x)\, ,{\Psi}_{\xi}(x)\right]=0$, 
we even need not to symmetrize or to pass the operator product in the 
r.h.s. of the HEqs (\ref{2.4}) into the normal-ordered form, but the 
careful definition of the Heisenberg ``currents'' $V_{-\xi(\Psi)}(x)$
in Eq. (\ref{2.5}) is still necessary. On the other hand, in order
to linearize Eq. (\ref {2.4}), we should first determine the
dynamics of these ``currents''. For this purpose we rewrite them
in a more convenient abstract form: $\nu= A$, 
\begin {gather}
\!\!\!\!\!\!\!\!\!\!\!\!\! 
\gamma^0\gamma^\nu= O^A,\;\; \xi
2\pi\lambda\epsilon_{\mu\nu}{\gamma}^0{\gamma}^\mu
=\tilde{O}^A_{-\xi},\;\; V_{-\xi(\Psi)}(x)\equiv \sum_A
\left(\Psi^\dagger_{-\xi}(x)O^A\Psi_{-\xi}(x)\right)\tilde{O}^A_{-\xi},
\label {2.6}
\end {gather}
and then by means of Eqs. (\ref{2.1})--(\ref{2.6}), we find for
these ``currents'' the formal HEqs
\begin {subequations}
\label {2.7_}
\begin {gather}
i\partial_0 V_{-\xi(\Psi)}(x)=
\left[V_{-\xi(\Psi)}(x)\,,H_{0(\Psi)}(t)\right]+
\left[V_{-\xi(\Psi)}(x)\,,H_{I(\Psi)}(t)\right]
\label {2.7_a} \\
= \sum_A\;\Biggl\{ \left(\Psi^\dagger_{-\xi}(x)\left[O^A
E_{-\xi}(P^1)-E_{-\xi}(P^{1\dagger})
O^A\right]\Psi_{-\xi}(x)\right)
\label {2.7_b} \\
+\left(\Psi^{\dagger}_{-\xi}(x)\left[O^AV_{\xi(\Psi)}(x)-
V^\dagger_{\xi(\Psi)}(x)O^A\right]\Psi_{-\xi}(x)\right)\Biggr\}
\left(\tilde O^A_{-\xi}\right),
\label {2.7_c} \\
\mbox {which, since }\;
O^A\propto \tilde{O}^B_{\xi}\propto I,\gamma^5,\;\;\,
\sum_A\sum_B\left({O}^B\right)\left(\left[{O}^A\,,
\tilde{O}^B_\xi\right]\right)\left(\tilde{O}^A_{-\xi}\right)= 0,
\label {2.7_d} \\
\mbox {read as}\;\; i\partial_0 V_{-\xi(\Psi)}(x)\equiv
\left[V_{-\xi(\Psi)}(x)\,,H_{(\Psi)}(t)\right]=
\left[V_{-\xi(\Psi)}(x)\,,H_{0(\Psi)}(t)\right] . 
\label {2.7_f}
\end {gather}
\end {subequations}
Of course the conditions (\ref{2.7_d}) are satisfied by any set of
the mutually commuting matrices ${O}^A,\tilde{O}^B$. But the final
Eq. (\ref{2.7_f}) implies also the absence of possible contributions 
from the Schwinger terms \cite {Singh}. Only under the assumption, as 
discussed in the Summary, the evolution of operators 
${V}_{-\xi(\Psi)}(x)$ is governed by the free one-particle Hamiltonian 
$H_{0(\Psi)}(t)$ only leaving just the line (\ref{2.7_b}). 
A simple main observation of our approach developed in \cite {korentanae}, 
\cite {vallkorenlevia}, is, that since the contribution responsible for the 
interaction vanishes, these Heisenberg operators (in the Heisenberg-field 
representation), as solutions of HEqs (\ref{2.7_f}) in a weak sense of DM 
are equivalent to itself in some free-trial-physical-field 
representation\footnote{That in fact takes place for all known solvable fermionic 
models.} which, according to the explicit form of (\ref{2.7_a}) given below by 
(\ref{B.1}), (\ref{B.2}), we should choose as free fields $\psi_{-\xi}(x)$ with 
the same mass $m_{-\xi}$ from Hamiltonian (\ref{2.2}). In other words, from 
homogeneous 
HEqs (\ref{2.7_f}) with (\ref{2.4}), (\ref{2.5}) we conclude that the DM 
$\Upsilon[\psi(x)]$ leaves the form of these ``currents'' invariant up to 
possible multiplicative constant fixed by initial condition at $t=0$ equal to 
unity. That is 
\begin {gather}
V_{-\xi(\Psi)}(x)=e^{iH_{(\Psi)}t}V_{-\xi(\Psi)}(x^1,0)e^{-iH_{(\Psi)}t}
= e^{iH_{0(\Psi)}t}V_{-\xi(\Psi)}(x^1,0)e^{-iH_{0(\Psi)}t}\longleftrightarrow
\label {2.8} \\
\longleftrightarrow e^{iH_{0(\psi)}t}V_{-\xi(\psi)}(x^1,0)e^{-iH_{0(\psi)}t}=
V_{-\xi(\psi)}(x),\;\mbox { or }\; 
V_{-\xi(\Upsilon[\psi])}(x)\stackrel{\rm w}{=}V_{-\xi(\psi)}(x), 
\nonumber
\end {gather}
where the last line takes place due to the fact, that \textit{just these and 
only these} ``currents'' (\ref{2.5}) determine the interaction in the HEqs 
(\ref{2.4}), so that the latter get a \textit{linearized} form
\begin {equation}
\partial_0{\Psi}_\xi(x)=-i\left[E_\xi(P^1)+V_{-\xi(\psi)}(x)\right]\Psi_\xi(x).
\label {2.9}
\end {equation}
Strictly speaking the operator product in the r.h.s. should now be symmetrized or 
normal-ordered. We show below, that the correct result however may be obtained 
proceeding from Eq. (\ref{2.9}), and argue the only possible weak sense of 
relation (\ref{2.8}). 

From a dynamical point of view the suggested linearization
condition (\ref{2.7_f}), as \textit {a free} equation of motion
for Heisenberg ``currents'', is a natural and straightforward
extension of the usual vector current conservation condition.
Indeed, supposing for a moment the unit matrices in Eq.
(\ref{2.6}) for all $A$, $O^A= I$, we immediately turn Eq.
(\ref{2.7_f}) into the well-known divergence identity for the
vector current, that is
\begin {gather}
i\partial_0 J^0_{-\xi(\Psi)}(x)=
\left[J^0_{-\xi(\Psi)}(x)\,,H_{0(\Psi)}(t)\right],\;\; \mbox { means }\;\;
\partial_\mu J^\mu_{-\xi(\Psi)}(x)=0,
\label {2.10} \\
\mbox {and }\;\;
\left[Q_{-\xi(\Psi)}(t)\,,H_{(\Psi)}(t)\right]=0,
\;\;\mbox{where }\;\;
Q_{-\xi(\Psi)}(t)=\int\limits^\infty_{-\infty}dx^1J^0_{-\xi(\Psi)}(x^1,t).
\label {2.11}
\end {gather}
Thus the conservation of the local Heisenberg vector current with
corresponding fermionic charge $Q_{-\xi(\Psi)}(t)$ is equivalent
to the fact, that the time evolution of its zero component
$J^0_{-\xi(\Psi)}(x)$ is governed by the free part
$H_{0(\Psi)}(t)$ of the total Hamiltonian $H_{(\Psi)}(t)$ only (a
similar phenomenon in the SG model is described in Ref. \cite
{changrajar}). However, since the component $J^0_{-\xi(\Psi)}(x)$
alone does not define entirely the r.h.s. of the equation of
motion (\ref{2.4}), then the conservation condition (\ref{2.10})
alone does not enough for the second line in Eq. (\ref{2.8}) to
exist, whereas the condition (\ref{2.7_f}) in common with Eq.
(\ref{2.4}) is enough. 

\section{Step 2. Functional integral representation of the solution}

The linearized HEqs (\ref{2.9}) allow to write their formal solution 
in the chosen representation of the physical fields (\ref{2.8}), as 
a time-ordered $({\cal T}$-ordered) exponential of different sets of 
mutually non-commuting operators, such as 
$X^1=x^1$ and $P^1=-i\partial_1,$ $\gamma^\nu$, $\psi_{-\xi}(x)$
\begin {gather}
\Psi_\xi(x^1,T)={\cal T}\left[\exp\left\{-i\int\limits_0^T
d{\eta}\left(E_\xi(P^1)+V_{-\xi(\psi)}(x)\right)\right\}\right]
\Psi_{\xi}(x^1,0)\, ,
\label {3.1}
\end {gather}
where the Schr{\"o}dinger initial conditions at $t=0$ are imposed.
Following Feynman \cite {feynmhibbs}, the Heisenberg field
(\ref{3.1}) for $T>0$ may be reconstructed from the initial field
$\Psi_{\xi}(x^1,0)$ using the matrix element of the evolution
kernel ${\cal Y}_{(\psi_{-\xi})}$, which in turn is expressed as a
functional integral over the phase space variables $x^1(\eta)$ and
$p^1(\eta)$
\begin{subequations} \label{3.2_}
\begin{gather}
\!\!\!
\Psi_{\xi}(x^1,T)=\int\limits^\infty_{-\infty}dy^1
{\cal Y}_{(\psi_{-\xi})}(T,x^1|y^1,0)\Psi_{\xi}(y^1,0)\,,
\label{3.2_a} \\
\!\!\!
{\cal Y}_{(\psi_{-\xi})}(T,x^1|y^1,0)=
\langle x^1|{\cal T}\left[\exp\left\{-i\int\limits_0^T d\eta\left(
E_\xi(P^1)+V_{-\xi(\psi)}(X^1,\eta)\right)\right\}\right]\!|y^1\rangle
\label{3.2_b} \\
\!\!\!
=\int\limits_{x^1(0)=y^1}^{x^1(T)=x^1}{\cal D}x^1(\eta)\int{\cal D}p^1(\eta)
\exp\left(i\int\limits_{0}^{T}d\eta\, p^1(\eta)\,{\dot x}^1(\eta)\right)
\nonumber \\
\cdot\,{\cal T}_{\gamma,\psi_{-\xi} }
\left[\exp\left\{-i\int\limits_{0}^{T}d\eta\left(E_\xi(p^1(\eta))+
V_{-\xi(\psi)}\left(x^1(\eta),\eta\right)\right)\right\}\right].
\label{3.2_c}
\end{gather}
\end{subequations}
However, the integrand still contains a ``mixture'' of different
structures, the $\gamma$-matrix and the field ones 
${\cal T}_{\gamma,\psi_{-\xi}}[\dots]$. Nevertheless, the explicit form 
(\ref{2.5}) of the free ``currents'' $V_{-\xi(\psi)}(x)=
\xi 2\pi\lambda\left[\gamma^5\,J_{-\xi(\psi)}^0(x)-I\,
J_{-\xi(\psi)}^1(x)\right]$ prompts the shift of functional
variable $p^1(\eta)=\underline{p}^1(\eta)-
\xi2\pi\lambda{J}^0_{-\xi(\psi)}\left(x^1(\eta),\eta\right)$ to
obtain for the matrix element, returning to the previous
designation $\underline{p}^1(\eta)\mapsto p^1(\eta)$ with
$\dot{x}^\mu(\eta)=(1,\dot{x}^1(\eta))$, the following expression
\begin{gather}
\!\!\!\!\!\!\!\!\!\!\!\!
{\cal Y}_{(\psi_{-\xi})}(T,x^1|y^1,0)=\!\!\!\!\!\!
\int\limits_{x^1(0)=y^1}^{x^1(T)=x^1}\!\!\!\!\!\!{\cal D}x^1(\eta)
{\cal T}_{\psi_{-\xi}}\left[\exp\left\{\!-i\xi2\pi\lambda\!
\int\limits_{0}^{T}\!\!d{\eta}
\epsilon_{\mu\nu}\dot x^\mu(\eta)J^\nu_{-\xi(\psi)}\left(x^1(\eta),\eta\right)
\right\}\right]
\nonumber \\
\!\!\!\!\!\!\!\!\!\!\!\!
\cdot\int{\cal D}p^1(\eta)
{\cal T}_\gamma\left[\exp\left\{-i\int\limits_{0}^{T}d\eta
\left(\gamma^5p^1(\eta)+\gamma^0m_\xi-p^1(\eta)\dot x^1(\eta)\right)\right\}
\right]\,.
\label{3.3}
\end{gather}
The most important property of this expression is the splitting of the
$\gamma$-matrix and the field  operator orderings, that moreover, are 
integrated over different functional variables. It is a simple matter to 
see that up to this point the space-time dimension does not play any 
essential role: if the linearization conditions (\ref{2.7_d}), (\ref{2.7_f}) 
are satisfied in higher-dimensional cases as well the above transformations 
yield results \cite {korentanae}, similar to Eq. (\ref{3.3}).

\section{Step 3. Bosonization and ${\cal T}$-ordering}

\subsection{Bosonization}

The local vector current of the free massive Dirac field is defined by
means of antisymmetrization or normal-ordering of a formal field
product \cite {bjorkdrell}. Due to the particular property of
two-dimensional space-time, like $\gamma^5\gamma^\mu=\epsilon^{\mu\nu}\gamma_\nu$ 
and the conservation law similar to (\ref{2.10}) these \textit {free} massive 
currents bosonize to pseudo-scalar fields $\Phi_{-\xi}(x)$ 
\cite {schrotruonweisz1}, 
\cite {whigh}
\begin{equation}
\epsilon_{\mu\nu}J^\nu_{-\xi(\psi)}(x)=
\frac{\partial_\mu\Phi_{-\xi}(x)}{\sqrt{\pi}},\;\mbox{ where }\;
J_{-\xi(\psi)}^\mu(x)=:\overline{\psi}_{-\xi}(x)\gamma^\mu\psi_{-\xi}(x):\, ,
\label{4.2}
\end{equation}
leading to the simple full-derivative representation
\begin{equation}
-\epsilon_{\mu\nu}\dot{x}^\mu(\eta)J^\nu_{-\xi(\psi)}\left(x^1(\eta),\eta\right)
=-\,\frac {1}{\sqrt{\pi}}\frac {d}{d\eta}\,
{\Phi}_{-\xi}\left(x^1(\eta),\eta\right),
\label{4.3}
\end{equation}
that recasts the first of the time-ordered exponentials in Eq.
(\ref{3.3}) to the form
\begin{equation}
{\cal T}_{\psi_{-\xi}} \Biggl[\ldots \Biggr]=
{\cal T}_{\Phi_{-\xi}}\left[\exp\left\{-ig
\int\limits_0^T d\eta \frac {d}{d\eta}\;
\Phi_{-\xi}\left(x^1(\eta),\eta\right)\right\}\right]\,,\;
\mbox { with }\; g=2\xi\lambda\sqrt{\pi}.
\label{4.4}
\end{equation}
Thus, the problem of ${\cal T}$-ordering of the exponential bilinear in 
the fermion field is transformed to the ${\cal T}$-ordering of an 
exponential linear in the pseudo-scalar field $\Phi_{-\xi}(x)$. This is a 
considerably more simple exercise at least for the free boson case, as is 
shown in Appendix. The necessary relation between the pseudo-scalar fields 
$\Phi_{-\xi}(x)$ and free pseudo-scalar fields $\phi_{-\xi}(x)$, turning the 
problem to the free-field case, is found in the next subsection.

\subsection{Reduction to the free massless pseudo-scalar fields}

A simplest way to relate the field $\Phi_{-\xi}(x)$ to the free field 
$\phi_{-\xi}(x)$ can be derived from the relation between the free massive 
$\psi_{-\xi}(x^1,t)$ and the free massless $\chi_{-\xi}(x^1,t)$ Dirac 
fields, with the formal unitary transformation operator $G_{-\xi}(t)$ 
\cite {blaso}, so that for arbitrary $t$, from
\begin {subequations} \label {4.5_}
\begin {gather}
i\gamma^\mu\partial_\mu\psi_{-\xi}(x)=m_{-\xi}\psi(x),\qquad
i\gamma^\mu\partial_\mu\chi_{-\xi}(x)=0,
\label {4.5_a} \\
\mbox {one has }\;\;
\psi_{-\xi}(x^1,t)={G}_{-\xi}^{-1}(t)\chi_{-\xi}(x^1,t)G_{-\xi}(t),
\label {4.5_b} \\
\mbox {as well as }\;\,
J^\mu_{-\xi(\psi)}(x^1,t)=
G^{-1}_{-\xi}(t)j^\mu_{-\xi(\chi)}(x^1,t)G_{-\xi}(t),
\label {4.5_c} \\
\mbox {where }\;\,
j_{-\xi(\chi)}^\mu(x)=:\overline{\chi}_{-\xi}(x)\gamma^\mu\chi_{-\xi}(x):.
\label {4.5_d}
\end {gather}
\end {subequations}
Now we temporarily omit for brevity the index $-\xi$, using
$m_{-\xi}=m$ and so on, to obtain by means of these Dirac
equations the following relations
\begin {subequations} \label {4.6_}
\begin {gather}
[R(t),\chi(x^1,t)]=im\gamma^0{\chi}(x^1,t),\;\;\mbox { for }\;\;
R(t)\equiv \left\{\partial_0 G(t)\right\}G^{-1}(t),
\label {4.6_a} \\
\mbox {so, that}\;\;
R(t)=-im\int\limits_{-\infty}^{\infty}d{y^1}\left(
\overline{\chi}(y^1,t)\chi(y^1,t)\right)=-R^{\dagger}(t),
\label {4.6_b} \\
G(t)={\cal T}_{\chi}\left[\exp\left\{\int\limits_0^td\eta R(\eta)\right\}\right]
G(0), \quad G^{\dagger}(t)=G^{-1}(t),
\label {4.6_c}
\end {gather}
\end {subequations}
where the solution (\ref{4.6_b}), (\ref{4.6_c}) of Eq. (\ref{4.6_a}) in terms 
of operators $\chi(x^1,t)$  is obtained by means of the corresponding
equal-time anti-commutation relations of the type (\ref{2.3}).

The following comments are in order. First of all, we note, that
an operator of the same type as $R(t)$ is present in the
Hamiltonian (\ref{2.2}) and should exist under appropriate
regularization for example in normal-ordered form \cite
{faberivano}, \cite {whigh}. The $G(t)$-transformation relates the
time evolutions of two different free quantized fields, each of
them is defined on a corresponding space of states. In fact, it is
a Bogoliubov transformation diagonalizing the free massive fermionic 
Hamiltonian in terms of the free massless fermion fields for arbitrary time 
$t$ similarly to \cite {blaso}.
The operator $G(0)$ can be used to impose additional initial conditions. 
Because this operator does not change the final results of further calculations, 
we eliminate it further on choosing for simplicity $G(0)=I$.

For the \textit {free massless} fermion current (\ref{4.5_d}) the above arguments,  
originative earlier the bosonization rules (\ref{4.2}), introduce now the \textit 
{free massless} pseudo-scalar field $\phi(x)$ \cite {gomessilva1}
\begin {equation}
\epsilon_{\mu\nu}j^\nu_{(\chi)}(x)=\frac {1}{\sqrt{\pi}}\,\partial_\mu\phi(x)\,,
\qquad \overline{\chi}(x)\chi(x)=\upsilon\cos\left(\beta\phi(x)\right),\qquad
\beta=2\sqrt{\pi},
\label {4.7}
\end {equation}
connecting this current and the field by the equation similar to Eq. (\ref{4.3}), 
so that both the equations amount to the relations between the pseudo-scalar 
fields only
\begin {gather}
\!\!\!\!\!\!\!\!\!\!\!\!
\frac {d\Phi\left(x^1(\eta),\eta\right) }{d\eta}=G^{-1}(\eta)
\left(\frac {d\phi\left(x^1(\eta),\eta\right)}{d\eta}\right)G(\eta),
\quad \Phi(x^1,t)=G^{-1}(t)\phi(x^1,t)G(t), 
\label {4.8} \\
\!\!\!\!\!\!\!\!\!\!\!\!
\mbox {where }\;
G(t)={\cal T}_{\phi}\left[\exp\left\{\int\limits_0^td\eta R(\eta)\right\}\right],
\quad R(\eta)=-im\upsilon\int\limits_{-\infty}^\infty
dy^1\cos(\beta\phi(y^1,\eta)),
\label {4.9}
\end {gather}
and the index $-\xi$ is assumed for all quantities. The value of
$\upsilon$ depends on the renormalization prescription chosen for
the normal-ordering procedures of boson and fermion fields \cite
{faberivano}, \cite {faberivano2} assumed in both sides of Eqs. (\ref{4.7}). 
These procedures are however irrelevant for our further operator
transformations, so we do not specify further the value of $\upsilon$.

The next comment concerns the properties of the field $\Phi(x)$.
These may be easily derived substituting into the free field equation 
for the field $\phi(x)$ its formal expression from Eq. (\ref{4.8}). Namely, 
for $x^0=t$
\begin {equation}
\partial_\mu\partial^\mu\phi(x)\equiv
\left(\partial^{2}_0-\partial^{2}_{1}\right)\phi(x^1,t)=0,
\;\;\mbox { where }\;\;\phi (x^1,t)=G(t)\Phi (x^1,t)G^{-1}(t).
\label {4._10}
\end {equation}
Using here Eqs. (\ref{4.9}) we arrive at the following form of
the second-time-derivative
\begin {equation}
\partial^2_0 \phi(x^1,t)=
G(t)[\partial^{2}_0\Phi(x^1,t)]G^{-1}(t)-[\partial_0\phi(x^1,t),R (t)].
\label {4._11}
\end {equation}
The commutator is easy computed by means of (\ref{4.9}) and the canonical 
commutation relation for the free pseudo-scalar field 
$\left[\partial_0\phi(x^1,t)\,,\phi(y^1,t))\right]=-i\delta(x^1-y^1)$ as 
follows
\begin {equation}
\left[\partial_0\phi (x^1,t),R(t)\right]=
m\upsilon\beta\sin\left(\beta\phi(x^1,t)\right).
\label {4._12}
\end {equation}
Thus the second relation (\ref{4._10}) leads to the SG equation of motion for the 
field $\Phi(x)$
\begin {equation}
\partial_\mu\partial^\mu\phi(x^1,t)=G(t)\left[\partial_\mu\partial^\mu\Phi(x)-
m\upsilon\beta\sin\left(\beta\Phi(x)\right)\right]G^{-1}(t)=0,
\label {4._13}
\end {equation}
which is reproduced for $\beta =2\sqrt{\pi}$ by the well-known Lagrangian density 
\cite {colem}, \cite {juricsazdo}, \cite {mande}, \cite {schrotruonweisz1}
\begin {equation}
{\cal L}^{SG}(x)=\frac{1}{2}\partial_\mu\Phi (x)\,\partial^\mu\Phi(x)-
\upsilon m\left[\cos\left(\beta\Phi(x)\right)-1\right].
\label {4._14}
\end {equation}

\subsection {Conversion of the ${\cal T}$-exponential into the unordered form}

Now we are ready to get rid of the ${\cal T}$-ordering in Eq.
(\ref{4.4}). This problem is simplified by using the well-known
operator formula for the ${\cal T}$-exponentials \cite {nazayk}
\begin {gather}
{\cal U}_{A+B}(T)={\cal U}_{A}(T)\,{\cal U}_{C}(T),\;\;\mbox{ where }\;\;
{C}(\eta)=\Bigl({\cal U}_{A}(\eta)\Bigr)^{-1}B(\eta)\,{\cal U}_{A}(\eta),
\label {4.10} \\
\mbox{with }\;\;{\cal U}_{A+B}(T)={\cal T}\left[\exp\left\{\int\limits_0^Td\eta
\biggl(A(\eta)+B(\eta)\biggr)\right\}\right],\;\;\mbox{ and so on.}
\nonumber
\end {gather}
Identifying here $A(\eta)=R(\eta)$, $B(\eta)=
-ig\left(d\phi(x^1(\eta),\eta)/d\eta\right)$, by means of the first Eq.
(\ref{4.8}) and Eq. (\ref{4.9}), we transform the expression on the r.h.s. of 
Eq. (\ref{4.4}) as follows
\begin {equation}
G^{-1}(T){\cal T}_\phi\left[\exp\left\{\int\limits_0^Td\eta\left(
R(\eta) -ig\,\frac {d\phi(x^1(\eta),\eta)}{d\eta}\right)\right\}\right],\;
\mbox{ with }\; g=\xi\lambda\beta.
\label {4.11}
\end {equation}
According to Eqs. (\ref{A.9}), (\ref{A.10}) of Appendix, the
${\cal T}$-exponential (\ref{4.4}) for the case of a \textit
{free} (pseudo-) scalar field is independent of the trajectory
$x^1(\eta)$ and depends only on its end points
\begin {gather}
W(T)\equiv{\cal T}_\phi\left[\exp\left\{-ig\int\limits_0^T
d\eta\frac{d\phi(x^1(\eta),\eta)}{d\eta}\right\}\right]=
U(T)U^{-1}(0)\,,
\label {4.12} \\
\mbox{where }\;\;
U(T)=\exp\left\{-\,i\frac{g^2}{4}\theta\left[1-\left(\dot x^1(T)\right )^2
\right]\right\}\, \exp\left[-ig\phi(x^1(T),T)\right]\,,
\label {4.13}
\end {gather}
that allows to apply again the relation (\ref{4.10}), written now
from left-to-right, to the last ${\cal T}_\phi-$ exponential in
Eq. (\ref{4.11}), just interchanging the roles of operators only:
$B(\eta)=R(\eta)$, $A(\eta)=
-ig\left(d\phi(x^1(\eta),\eta)/d\eta\right)$, recasting Eq. (\ref{4.11}) into 
the form
\begin {gather}
G^{-1}(T)U(T){\cal T}_\phi\left[
\exp\left\{\int\limits_0^T d\eta\,U^{-1}(\eta)
R(\eta)U(\eta)\right\}\right] U^{-1}(0)\, .
\label {4.14}
\end {gather}
From the Eqs. (\ref{4.9}), (\ref{4.13}), with the locality condition for 
the field $\phi(x)$, originated in Eqs. (\ref{A.1}), (\ref{A.2}) of Appendix, 
it is a simple matter to see that, when $x=x(\eta)=(\eta,x^1(\eta))$, 
$y=(\eta,y^1)$, then $x(\eta)-y=(0, x^1(\eta)-y^1)$,   
\begin {gather}
[\phi\left(x(\eta)\right),\phi(y)]= -i\,D_0\left(x(\eta)-y\right)=0,\quad 
U^{-1}(\eta)R(\eta)U(\eta)= R(\eta), 
\label {4.15}
\end {gather}
and the integrand in Eq. (\ref{4.14}) is reduced to $R(\eta)$. Therefore, 
by means of (\ref{4.8}), we arrive at the final expression for the 
${\cal T}$-exponential of Eqs. (\ref{4.4}) = (\ref{4.14}), as
\begin {gather}
G^{-1}(T)U(T)G(T)U^{-1}(0)=
\Omega(T,0)\exp\left[-ig\Phi(x(T))\right]\exp\left[ig\Phi(x(0))\right],
\label {4.16} \\
\mbox{where }\;\,\Omega(T,0)=
\exp\left\{-\,i\frac{g^2}{4}\left(
\theta\left[1-\left(\dot x^1(T)\right)^2\right]-
\theta\left[1-\left(\dot x^1(0)\right)^2\right]\right)\right\}, 
\label {4.17}
\end {gather}
is a c-number phase factor. It different from unity, for example, if absolute 
value of the velocity $|\dot x^1(t)|$ during the evolution has crossed 
the speed of light an odd number of times.

\section {Step 4. Initial and Heisenberg fields}

Now we are finally able to obtain the solution of the Federbush
model \cite {castralvarfring}, \cite {whigh}. First of all we
notice that the above obtained expression (\ref{4.16}) for the
${\cal T}$-exponential is fully independent of the trajectory
$x^1(\eta)$ and factorizes out from the path integral in Eq.
(\ref{3.3}) which is then immediately calculated by definition
(\ref{3.2_a}), (\ref{3.2_c}) as free evolution kernel
$Y_{\xi}^{(0)}$, expressing the time evolution of the another free
fermion field with the mass $m_\xi$ 
\begin {subequations} \label {5.0_}
\begin {gather}
\!\!\!\!\!\!\!\!\!
{\cal Y}_{(\psi_{-\xi})}(T,x^1|y^1,0)=\Omega(T,0)
\exp\left[-ig\Phi_{-\xi}(x(T))\right]\exp\left[ig\Phi_{-\xi}(x(0))\right]
Y_\xi^{(0)}(T,x^1|y^1,0)\,,
\label {5.0_c} \\
\!\!\!\!\!\!\!\!\!
\mbox{where }\;\;
Y_\xi^{(0)}(T,x^1|y^1,0)=\theta(T)\exp\left[-iTE_\xi\left(P^1\right)
\right]\,\delta(x^1-y^1)
\label {5.0_b} \\
\!\!\!\!\!\!\!\!\!
=\int\limits_{x^1(0)=y^1}^{x^1(T)=x^1}{\cal D}x^1(\eta)
\int{\cal D}p^1(\eta){\cal T}_\gamma\left[
\exp\left\{-i\int\limits_{0}^{T}d\eta
\left(\gamma^5 p^1(\eta)+\gamma^0m_\xi-p^1(\eta)\dot x^1(\eta)\right)
\right\}\right]\,,
\nonumber \\
\!\!\!\!\!\!\!\!\!
\mbox{so, that }\;\;
\int\limits^{\infty}_{-\infty}dy^1\,
Y_\xi^{(0)}(T,x^1|y^1,0)\;\psi_{\xi}(y^1,0)=\psi_{\xi}(x^1,T)\,.
\label {5.0_a}
\end {gather}
\end {subequations}
In order to check Eq. (\ref{5.0_b}) we need only the following relation 
\cite {polya}
\begin {gather*}
\int\limits^{x^1(T)=x^1}_{x^1(0)=y^1}{\cal D}x^1(\eta)\,[\dots]=
\int{\cal D}{\dot x}^1(\eta)\delta\left(x^1-y^1-\int\limits_0^Td\eta\,
{\dot x}^1(\eta)\right)\,[\dots], 
\end {gather*}
and the Fourier-representation for the delta-function. From the Eqs.
(\ref{5.0_}) above it is a simple matter to see, that in order to obtain the 
correct solution for $x^1(T)=x^1$, the initial field for $x^1(0)=y^1$
should be chosen as follows
\begin {gather}
\Psi_{\xi}(y^1,0)=
\exp\left[-\,ig\Phi_{-\xi}(y^1,0)\right]\psi_{\xi}(y^1,0),
\quad \; g=2\xi\lambda\sqrt{\pi},
\label {5.1} \\
\mbox{so, that }\;\,
\Psi_{\xi}(x^1,T)\stackrel{\rm w}{=}\Upsilon [\Phi_{-\xi},\psi_\xi]
=\Omega(T,0)\exp\left[-\,ig\Phi_{-\xi}(x^1,T)\right]\psi_{\xi}(x^1,T).
\label {5.2}
\end {gather}
It should be stressed that, as a solution of the \textit {linear 
homogeneous equation} (\ref{2.9}), this expression may be easily
renormalized following Wightman \cite {schrotruonweisz1}, \cite {whigh} by 
dividing the r.h.s. of Eq. (\ref{5.2}) on the vacuum expectation value 
$\langle 0|\exp\left[-\,ig\Phi_{-\xi}(x))\right]|0\rangle$. The factor 
$\Omega(T,0)$ (\ref{4.17}) is generated in Eq. (\ref{A.8}) of Appendix
by non-vanishing contribution of zero modes into the dimensionless 
Pauli-Jordan commutator function (\ref{A.2}) of the two-dimensional free 
massless (pseudo-) scalar fields. As was shown by Faber et al. 
\cite {faberivano}, \cite {faberivano2}, these collective modes are not 
affected by the dynamics and their consecutive removal makes meaningful 
the theory of the free massless two-dimensional (pseudo-) scalar field. 
With this interpretation the value $|\dot x^1(t)|$ is the constant velocity 
of the ``center of mass system'' of these zero modes decoupled from everything, 
and $\Omega(T,0)\equiv 1$ for all $T$. 

\section{Summary}

We have shown, how the solution of the Federbush model may be
obtained by direct integration of the linearized formal HEqs 
using the notion of physical fields. These fields are conventionally taking 
as the Schr{\"o}dinger ones, because the latter as well as Heisenberg fields  
form a complete set of fields itself, unlike the asymptotic \textit {in (out)} 
fields \cite {umeza}, what is also useful for construction of explicit solutions 
of the bound-state problem \cite {korentanae}, \cite{vallkorenlevia}. 
We have revealed the fundamental 
role of the free massless pseudo-scalar field for this integration procedure. 
We have found, that the solvability of the Federbush model, besides the properties 
of the linearization (\ref{2.8}) and the bosonization (\ref{4.2}), hides in the 
full-derivative expression (\ref{4.3}) for the interaction term with an arbitrary 
trajectory in the time-ordered exponential of the path integral representation for
the evolution kernel (\ref{3.3}) of the Heisenberg field.
It is worth to note that the method suggested here may be applied to another known 
solvable two-dimension models like the Thirring and derivative coupling models   
\cite {gomessilva1}, the model in Ref. \cite {belveamara}, as well as to some 
four-dimensional ones \cite {korentanae}.

Our last comments concern a possible contribution to Eq. (\ref{2.7_a}) 
from the Schwinger terms. For the massless Thirring model it may be 
shown that their invariant contribution vanishes independently of the value 
of the constant in the current commutator. On the one hand, it is 
well-known \cite {faberivano}, \cite {vladimirov}, that this constant directly 
depends on the chosen physical field representation -- the chosen space of 
states. On the other hand, it was shown \cite {Singh}, that for the 
Federbush model its value strongly depends on the chosen point-splitting 
prescription for the \textit {Heisenberg vector current} in the
HEqs  (\ref{2.4}), (\ref{2.5}), (\ref{2.7_}). That is why in this
work we prefer to deal with local currents only. 

Indeed, for these currents, because of $\gamma^1=\gamma^0\gamma^5$, the 
``non-interacting'' and ``interacting'' parts of the HEqs
(\ref{2.7_a})--(\ref{2.7_c}) may be transcribed straightforwardly
as follows
\begin {gather}
i\partial_0 V_{-\xi(\Psi)}(x)-\left[V_{-\xi(\Psi)}(x)\,,H_{0(\Psi)}(t)
\right]
=\gamma^5\,g\sqrt{\pi}\biggl\{i\partial_\nu\left(\overline{\Psi}_{-\xi}(x)
\gamma^{\nu}\Psi_{-\xi}(x)\right)\biggr\}
\label {B.1} \\
+I\,g\sqrt{\pi}\biggl\{i\partial_\nu\left(\overline{\Psi}_{-\xi}(x)
\gamma^5\gamma^{\nu}\Psi_{-\xi}(x)\right)-
2m_{-\xi}\left(\overline{\Psi}_{-\xi}(x)\gamma^5\Psi_{-\xi}(x)\right)\biggr\},
\nonumber \\
\left[V_{-\xi(\Psi)}(x)\,,H_{I(\Psi)}(t)\right]=\gamma^5\,g\sqrt{\pi}
\left[J^0_{-\xi(\Psi)}(x)\,,H_{I(\Psi)}(t)\right]
\label {B.2} \\
-I\,g\sqrt{\pi}\left[J^1_{-\xi(\Psi)}(x)\,,H_{I(\Psi)}(t)\right].
\nonumber
\end {gather}
Matching the independent matrix structures $I$ and $\gamma^5$ of
these two equations we easily recognize both vector current and
axial current divergence identities if the ``interacting''
part (\ref{B.2}) vanishes identically. Therefore the conservation of the
Heisenberg vector current (\ref{2.10}) preserves here the absence
of the full contribution (\ref{B.2}) from the Schwinger terms in the r.h.s. of 
Eqs. (\ref{2.7_a})--(\ref{2.7_c}), because for the given interaction 
$H_{I(\Psi)}(t)$ (\ref{2.1}), (\ref{2.2}) both commutators in the r.h.s. of 
Eq. (\ref{B.2}) reduce to the one and the same Schwinger term in the one and the 
same commutator 
$\left[J^1_{-\xi(\Psi)}(x)\,,J^0_{-\xi(\Psi)}(y)\right]\Bigr|_{x_0=y_0}$, and 
are vanishing or not only both together. It means that the bosonization rules 
like (\ref{4.2}), (\ref{4.7}) cannot be applied here to the local Heisenberg 
vector current $J^\nu_{-\xi(\Psi)}(x)$ directly, unlike \cite {colem}, 
\cite {faberivano}, \cite {gomessilva1}.

The observed contradiction may be overcomed following Ref. \cite {sokolov} 
in QED, if we impose bosonization rules to the \textit {free} fermion currents 
(\ref{4.2}), (\ref{4.7}) \textit {only}, and if \textit {only} 
weak correspondence (\ref{2.8}) between the Heisenberg and free currents take 
place. Thus the Schwinger term does not contribute into the HEqs (\ref{2.7_a}), 
because it becomes meaningful 
\textit {only after} the identification transformation (\ref{2.8}) 
determining the representation space. These arguments, besides of all, 
demonstrate a self-consistence of the suggested linearization procedure   
(\ref{2.4}), (\ref{2.5}) $\mapsto$ (\ref{2.7_f}) $\mapsto$ (\ref{2.8}), 
(\ref{2.9}) with obtained solution (\ref{5.2}) or (\ref{B.3}), (\ref{B.4}). 

Thus we come to the conclusion: the transformation of the ``currents'' 
(\ref{2.8}), generated for the Federbush model by the DM (\ref{5.2})
with the pseudo-scalar SG field $\Phi_{-\xi}(x^1,t)$ and normal-ordered 
with respect to the Schr{\"o}dinger free-fermion-physical fields 
$\psi_{\pm\xi}(x)$ 
\begin {gather}
\Psi_{\xi}(x^1,t)\stackrel{\rm w}{=}\Upsilon[\psi_{\pm\xi}]=
:\exp\left[-2i\xi\lambda\sqrt{\pi}\,\Phi_{-\xi}(x^1,t)\right]\psi_{\xi}(x^1,t):,
\label {B.3} \\
\mbox {where }\;\;
\Phi_{-\xi}(x^1,t)=\frac{\sqrt{\pi}}{2}\left(
\int\limits^{x^1}_{-\infty}dy^1-\int\limits_{x^1}^{\infty}dy^1\right)
J^0_{-\xi(\psi)}(y^1,t),
\label {B.4}
\end {gather}
(see (\ref{4.2})) 
leaving invariant the local form of the both components of vector current
$J^\nu_{-\xi(\Psi)}(x)\stackrel{\rm w}{=}J^\nu_{-\xi(\psi)}(x)$, as well as 
conservation law (\ref{2.10}), changes their algebra, extending it by 
the Schwinger term, whereas for the conserving charges (\ref{2.11})
\begin {gather}
Q_{-\xi(\Psi)}(t)\stackrel {\rm w}{=}Q_{-\xi(\psi)}(t)=\frac
{\Phi_{-\xi}(\infty,t)-\Phi_{-\xi}(-\infty,t)}{\sqrt{\pi}}\equiv
\frac 2{\sqrt{\pi}}\,\Phi_{-\xi}(\infty,t), \label {B.5}
\end {gather}
the algebra remains unchanged. 

We  are grateful to Prof. Vall A.N. and Prof. Leviant V.M. for useful
discussions, and to the referee of JNMP for suggesting improvements of the 
manuscript.

\section {Appendix}

Let us consider more general time-ordered exponential of the total 
derivative of the free \textit {massive} (pseudo) scalar field 
$\varphi(\eta)=g\phi(x^1(\eta),\eta)$, taken on an arbitrary trajectory 
$x^1(\eta)$, where its commutator function $\Delta_m(t|\eta)$ is expressed 
via the Bessel function ${\cal J}_0(z)$ as
\begin {gather}
\left[\varphi(t),\varphi(\eta)\right ]=\Delta_m(t|\eta)=
-i g^2 D_m\left(x(t)-x(\eta)\right),
\label {A.1} \\
D_m(x)=2\pi i\int\frac{d^2k}{(2\pi)^2}\,\varepsilon(k^0)
\delta\left(k^2-m^2\right)e^{-i(kx)}=\frac{\varepsilon(x_0)}{2}\,
\theta\left(x^2\right){\cal J}_0\left(m\sqrt{x^2}\right).
\label {A.2}
\end {gather}
Since this commutator is a c-number, by repeated application of
the well known formula:
$\exp(A)\exp(B)=\exp(A+B)\exp(\{1/2\}[A,B])$, one can obtain for
the ${\cal T}-$ exponential
\begin {gather}
\!\!\!\!\!\!\!\!\!
{\cal T}_{\varphi}\left[\exp\left\{-i\int\limits_{0}^{T}dt\frac{d\varphi(t)}
{dt}\right \}\right ]=\exp \left \{-i \varphi(T)\right \}\exp \left\{i\varphi(0)
\right\}\exp\left\{-\,\frac {1}{2}\Delta_m(T|0)\right\}
\label {A.3} \\
\!\!\!\!\!\!\!\!\! \cdot\,\exp\left [-\,\frac
{1}{2}\int\limits_{0}^{T}d\eta\int\limits_{0}^\eta
d\tau\left\{\frac
{d}{d\eta}\frac{d}{d\tau}\Delta_m(\eta|\tau)\right\} \right ],\;
\nonumber 
\end {gather}
and transcribes it, by means of the locality condition (\ref{4.15}) and the 
next equality 
\begin {gather}
\!\!\!\!\!\!\!\!\!
\int\limits_{0}^{T}d\eta \int\limits_{0}^{{\eta}}{d\tau}\left\{
\frac {d}{d{\eta}}\frac {d}{d\tau}\Delta_m(\eta |\tau)\right\}=
\Delta_m(T|T)-\Delta_m(T|0)-\int\limits_{0}^{T}d\eta\left[\frac{d}{d\tau}
\Delta_m(\eta|\tau)\biggr|_{\tau=\eta}\right],
\nonumber 
\end {gather}
as
\begin {gather}
\!\!\!\!\!\!\!\!\! 
{\cal T}_\varphi\Biggl[\ldots
\Biggr]= \exp\left \{-i \varphi(T)\right \}\exp\left \{i
\varphi(0)\right \}\exp\left
\{\frac{1}{2}\int\limits_{0}^{T}d\eta\left [\frac
{d}{d\tau}\Delta_m(\eta|\tau) \biggr|_{\tau=\eta}\right ]\right\}.
\label {A.4}
\end {gather}
The function $\Delta_m(\eta|\tau)$ is discontinuous on
the light-cone, at $\eta=\tau$, and its derivative has a
singularity there. Therefore the expression for the integral in
the last exponential is indefinite and should be redefined as the
following limit with $\sigma\to +0$ of
\begin {gather}
\int\limits_{0}^{T}d\eta\left[\frac{d}{d\tau}
\Delta_m(\eta|\tau)\biggr|_{\tau=\eta-\sigma}\right ]=
\frac{g^2}{i}\int\limits_{0}^{T}d\eta\; 2\pi i\int
\frac{d^2k}{(2\pi)^2} \varepsilon(k^0)\delta(k^2-m^2)
\label {A.5} \\
\cdot\,\left[\frac {d}{d\tau}e^{-i\left(k\cdot(x(\eta)-x(\tau))\right)}
\biggr|_{\tau=\eta-\sigma}\right].
\nonumber
\end {gather}
For the latter derivative, using the Taylor expansion of the
trajectory $x^1(\eta-\sigma)$ with respect to $\sigma>0$, and
changing the momentum variables as $ q^\mu=\sigma k^\mu $, we obtain
\begin {gather}
\frac {d}{d\tau}e^{-i\left
(k^0(\eta-\tau)-k^1(x^1(\eta)-x^1(\tau))\right )}\biggr
|_{\tau=\eta-\sigma}\!\!\! = \left [\frac{d}{d\eta}-\frac 1\sigma
\frac{\partial}{\partial\gamma}\right
]e^{-i\gamma\left(q^0-q^1\dot x^1(\eta)
\right)}\biggr|_{\gamma=1}. \label {A.5_}
\end {gather}
After these changes the integral (\ref{A.5}) becomes
\begin {gather}
\frac{g^2}{i}\int\limits_{0}^{T}{d\eta}\left[\frac{d}{d\eta}-\frac 1\sigma
\frac{\partial}{\partial\gamma}\right]2\pi i\int\frac{d^2q}{(2\pi)^2}
\varepsilon (q^0)\delta(q^2-\sigma^2 m^2)e^{-i\gamma\left(q^0-q^1\dot 
x^1(\eta)\right)}\biggr|_{\gamma=1}
\label {A.6} \\
=\frac{g^2}{2i}\int\limits_{0}^{T}\!{d\eta}\!
\left[\frac{d}{d\eta}-\frac 1\sigma\frac{\partial}{\partial\gamma}\right]
\!\varepsilon(\gamma)\theta\left(\!\gamma^2\!
\left[1-\left(\dot x^1(\eta)\right)^2\right]\right)\!{\cal J}_0\!\left(\sigma m
\gamma\sqrt{1-\left(\dot x^1(\eta)\right)^2}\right)\biggr|_{\gamma=1}.
\nonumber
\end {gather}
Because ${\cal J}_0(z)=1+O(z^2)$, it is easy to see, that the mass
dependance of (\ref{A.2}) leads here just to terms of order
$\sigma$. So, as $\sigma\rightarrow +0$, the direct substitution
$\delta(q^2-\sigma^2m^2)\mapsto\delta(q^2)$, gives for the
integral (\ref{A.6})
\begin {equation}
\frac{g^2}{2i}\int\limits_{0}^{T}{d\eta}\left [ \frac{d}{d\eta}-\frac 1\sigma
\frac{\partial}{\partial\gamma}\right ]\varepsilon(\gamma)\theta\left (\gamma^2
\left [1-\left (\dot x^1(\eta)\right )^2\right ]\right )\biggr|_{\gamma=1}.
\label {A.7}
\end {equation}
The integrand in (\ref{A.7}) does not depend on $\gamma$ at all
near the point $\gamma=1$, and its corresponding derivative
vanishes, eliminating the contribution singular at
$\sigma\rightarrow +0$. The remaining integral on $d\eta$ is taken
immediately, leading to the expression for the limit
\begin {equation}
\lim_{\sigma\rightarrow +0}\int\limits_{0}^{T}d\eta\left[\frac{d}{d\tau}\Delta_m
(\eta|\tau)\biggr|_{\tau=\eta-\sigma}\right ]=
\frac{g^2}{2i}\left\{\theta\left[1-\left(\dot x^1(T)\right)^2\right]-
\theta\left[1-\left(\dot x^1(0)\right)^2\right]\right\}.
\label {A.8}
\end {equation}
As a result, the time-ordered exponential (\ref{A.3}) takes the
following form
\begin {gather}
{\cal T}_{\varphi}\left [ \exp\left \{-i\int\limits_{0}^{T}dt\frac{d\varphi(t)
}{dt}\right \}\right ]=U(T)U^{-1}(0),\qquad \qquad
\label {A.9} \\
\mbox {where }\;\,
U(T)=\exp\left \{-\,i\frac{g^2}{4}\theta\left [1-\left (\dot x^1(T)\right )
^2\right]\right\}\exp\left\{-i\varphi(T)\right\}.
\label {A.10}
\end {gather}
This result is of fundamental importance for all computations,
because it depends on the values of an unknown trajectory
$x^1(\eta)$ and velocity $\dot x^1(\eta)$ only at the initial and
final points, $t=0$ and $t=T$. Moreover, the phase factor from 
(\ref{A.4}), (\ref{A.8}) in (\ref{A.9}) becomes the same for both
the massive and massless cases, that, on the one hand, corresponds
to mass independence of the light-cone behavior, but on the other
hand, due to the dimensionless nature of the commutator function
(\ref{A.2}), represents by the construction (\ref{A.5})--(\ref{A.7}) 
the non-trivial infrared properties of the free massless (pseudo-) scalar 
field, such as the existence and the non-vanishing contribution of its 
zero modes \cite {faberivano}, \cite {faberivano2}.

\label {lastpage}
\end {document}